# Developing of New Facets of Indirect Modeling in the Geosciences

Hamed.O.Ghaffari, Hadi.Shakeri and Mostafa.Sharifzadeh
*Department of Mining, Metallurgical and petroleum Engineering, Amirkabir University of Technology, Tehran, Iran*

**ABSTRACT:** In this paper, we describe some applications of Self Organizing feature map Neuro-Fuzzy Inference System (SONFIS) and Self Organizing feature map Rough Set (SORST) in analysis of permeability at a dam site and lost circulation in the drilling of three wells in Iran. Elicitation of the best rules on the information tables, exploration of the dominant structures on the behaviour of systems while they fall in to the balance of the second granulation level (rules) and highlighting of most effective attributes (parameters) on the selected systems, are some of the benefits of the proposed methods. In the other process, using complex networks (graphs) theory - as another method in not 1:1 modelling branch- mechanical behaviour of a rock joint has been investigated.

## 1. INTRODUCTION

Developing of not 1:1 mapping modeling in the several fields of engineering, has allocated a numerous researches which employs a large scope of logical, stochastic and natural computing methods. Regard to the several facets of uncertainties and complexity intrinsic of natural events especially in the geosciences fields, utilizing of indirect modeling in parallel to the direct analysis methods (1:1 mapping instruments) can emerge new supplementary information on the given problem. On the other hand, information resulted from the measurements, observations, experiences, education or/and collection of other information coherently are accompanied with ambiguity, raw, superficial (adjective attributes) and superfluous knowledge. In the other view, reduction of complexity is one of the main steps at the understanding of the complex systems behaviors. Complexity reduction of the system by its granulation, gives another opportunity to development of a human apprehensible model with less computational. Construction of granules and computation the granules are two main issues in this area. The former one deals with the representation and construction of granules, while the later is related with the employing of granules in problem solving and understanding human apprehensible model with less computational cost [1]. Upon this we develop different granulation methods based on intelligent systems and approximate reasoning methods. It must be noticed the route of granulation and degranulation (organizing of information) is in the current behavior of human cognition [2]. The main distinguished facets of the soft granules (related with indirect modeling) consist of: set theory, interval analysis, fuzzy set, rough set. Also, each of these theories considers part of uncertainty of information (data, words, pictures, etc).

In other view, possible relations between the building particles of complex systems can be revealed in complex networks, which describe a wide range of systems [3]. The success in the describing of interwoven systems –with nontrivial network structures- using physical tools as a major reductionism is associated with the simplifications of the interactions between the elements where there is not possible vagueness. Employing of statistical mechanics tools gives a good framework for analysis of these systems. Variations (growing or decaying) of the complex networks and also interactions between networks determine the future destination of the given system. As a complementary view to the logical describing of an information field (table), one may consider each inference machine or systems with inputs/output(s) as a network where interactions between attributes (conditional and decisional) or rules and elements are regulated in a way that the general frame shows best answer to the whole. In this study our aim is to developing and utilizing of granular computing methods in the large scale permeability assessment and mud lost circulation in the drilling of three wells in the Fars area –Iran. On this track, two new hybrid intelligent systems /approximated methods have been developed as if the hidden complexities within the information are captured with the approximated rules as a level of soft granules. In the separated section, based on complex networks, the mechanical behavior of a rock joint under shear displacement has been investigated. The exposed results show how phase transition step in the evolving of a surface can be reassessed with the lower costs (measurements) using possible assigned complex networks.

## 2. SONFIS &SORST

In this section based upon Self Organizing feature Map (SOM) [4], adaptive Neuro-Fuzzy Inference System (NFIS) [5] and Rough Set Theory (RST) [6], we reproduce: Self Organizing Neuro-Fuzzy Inference System (SONFIS) and Self Organizing Rough Set Theory (SORST) [7], [8]. The mentioned algorithms use four basic axioms upon the balancing of the successive granules assumption:

- Step (1): dividing the monitored data into groups of training ,validation, and testing data
- Step (2): first granulation (crisp) by SOM or other crisp granulation methods
  
  Step (2-1): selecting the level of granularity randomly or depend on the obtained error from the NFIS or RST (regular neuron growth)
  
  Step (2-2): construction of the granules (crisp).
- Step (3): second granulation (fuzzy or rough granules) by NFIS or RST
  
  Step (3-1): crisp granules as a new data.
  
  Step (3-2): selecting the level of granularity; (Error level, number of rules, strength threshold, scaling of inserted information...)
  
  Step (3-3): checking the suitability; Close-open iteration: referring to the real data and reinspect closed world
  
  Step (3-4): construction of fuzzy/rough granules.
- Step (4): extraction of knowledge rules

Balancing assumption is satisfied by the close-open iterations: this process is a guideline to balancing of crisp and sub fuzzy/rough granules by some random/regular selection of initial granules or other optimal structures and increment of supporting rules (fuzzy partitions or increasing of lower /upper approximations ), gradually.

The overall schematic of SONFIS and SORST has been shown in Figures 1, 2. In first granulation step, we can employ two situations: use a linear relation (or other functional regulation) is given by:

$$N_{t+1} = \alpha N_t + \Delta_t ; \Delta_t = \beta E_t + \gamma \quad , \qquad (1)$$

where $N_t = n_1 \times n_2 ; |n_1 - n_2| = Min.$ is number of neurons in SOM or Neuron Growth (NG); $E_t$ is the obtained error (measured error) from second granulation on the test data and coefficients must be determined, depend on the used data set. In second situation, the structure of SOM is identified, randomly. One may interpret Eq. (1) as evolution of crisp granules in time step t which are depend on the effects of the pervious state of granules ($\alpha$), the impact of the regulator (here NFIS or RST) performance ($\beta$) and other external forces ($\gamma$). Obviously, one can employ like manipulation in the rules (second granulation) generation part, i.e., number of the rules.

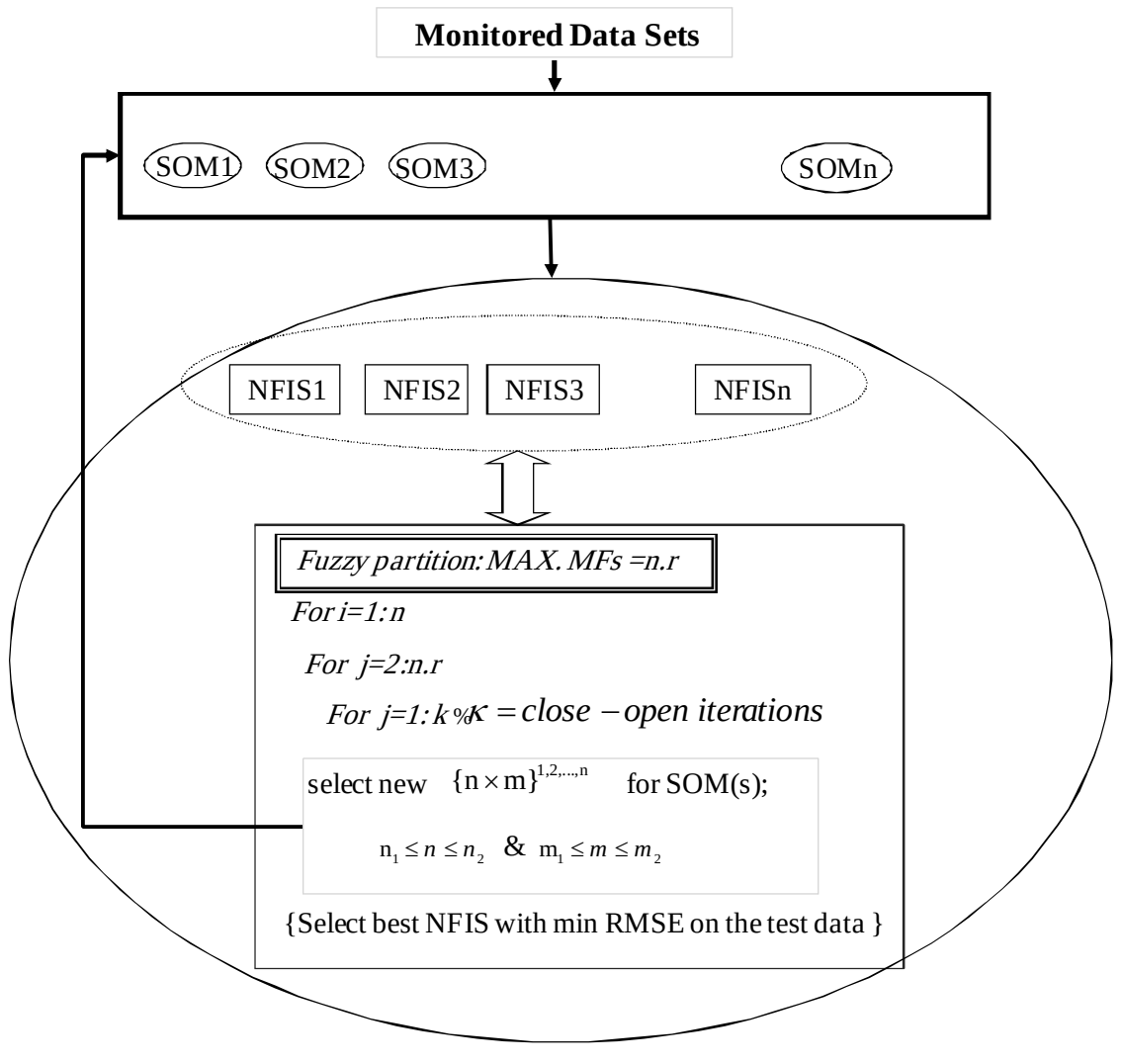

Fig.1.Self Organizing Neuro-Fuzzy Inference System (SONFIS)

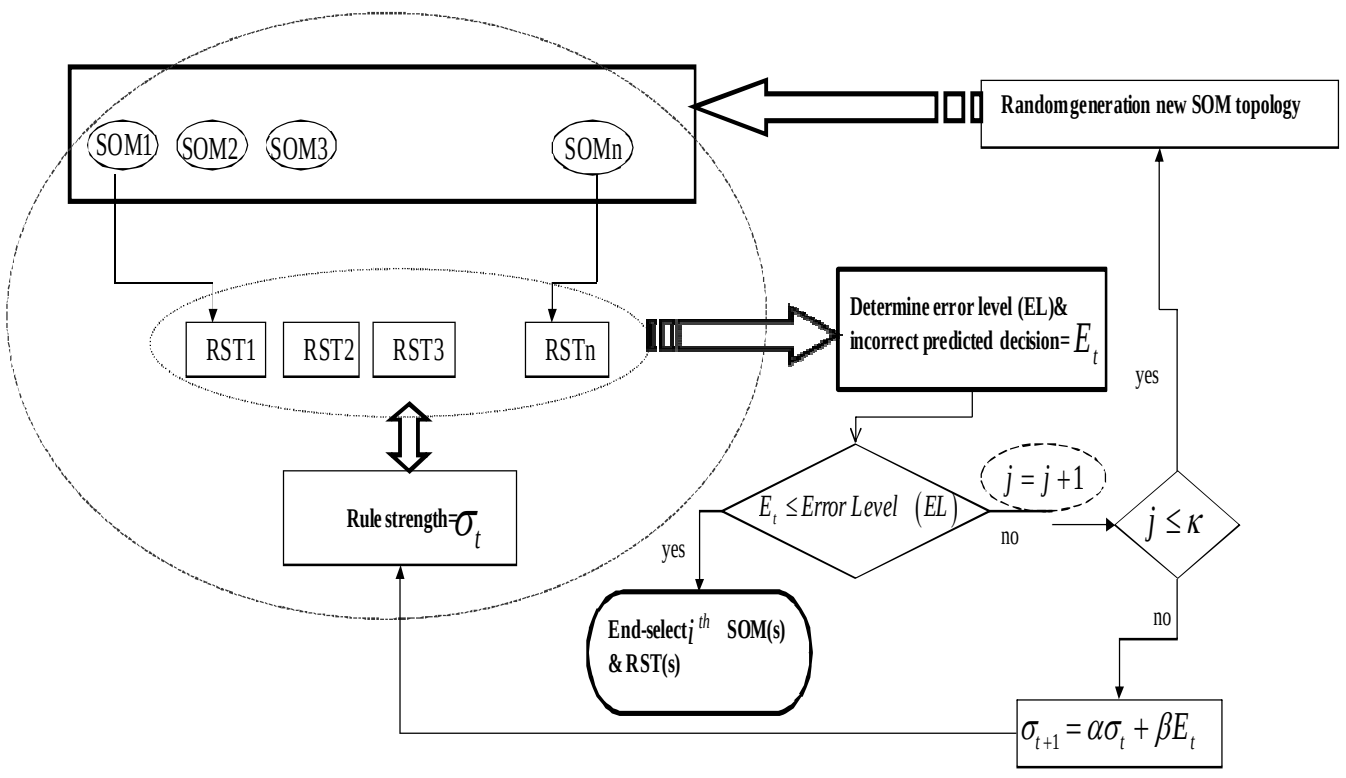

Fig.2.Self Organizing Rough SeT (SORST)

In second algorithm (figure 2), RST instead of NFIS has been proposed where regulation of rule generation in rough set is controlled by strength of each rule. Applying of SOM as a preprocessing step and discretization tool is second process. Categorization of attributes (inputs/outputs) is transferring of the attribute space to the symbolic appropriate attributes. In fact for continuous valued attributes, the feature space needs to be discretized for defining indiscernibilty relations and equivalence classes. We discretize each feature in to some levels by SOM, for example "low, medium, and high" for attribute "a". In this study, we use SONFIS and SORST to evaluation of mud loss in three wells at the Fars area and large scale permeability, respectively.

## 2.1. Lost circulation Analysis using SONFIS

Lost circulation is one of the oldest, most time consuming, and costly problems encountered in drilling a well [9]. During the last century, lost circulation (LC) has presented great challenges to the petroleum industry, causing great expenditures of cash and time to fighting the problem. Trouble costs for mud losses, wasted rig time, ineffective remediation materials and techniques, and in the worst cases—for lost holes, sidetracks, bypassed reserves, abandoned wells, relief wells, and lost petroleum reserves have continued into this century. The risk of drilling wells in areas known to contain these problematic formations is a key factor in decisions to approve or cancel exploration and development projects [10]. Generally, four types of formations are responsible for lost circulations: (1) natural or induced formation fractures, (2) vugular or cavernous formations, (3) highly permeable formations, and (4) unconsolidated formations. In this part of paper, we use a SONFIS structure with random regulation in first layer (granulation) on the extracted information table (figure 3a) at the three wells in Fars area. Fars area is located in the south of Iran with about 150 $Km^2$, in a geological view it is a part of Zagros sedimentary basin, properties of source rock is in such condition that we can't expect a lot of oil in place, but there are huge amount of gas in the various formations, especially in Dahrom and upper formations. During drilling of wells in this area major problem was loss of mud in to formations. To analysis of the mud loss patterns in this area, we collected 392 objects consist 12 attributes (inputs/output). Among these objects, 263, 80 and 52 patterns were selected as training, testing and validation data, respectively. Analysis of the accumulated data sets is started by setting the number of close-open iterations and maximum number of rules equal to 10 and 4 in SONFIS-R, respectively. The error measure criterion in SONFIS is Fourth Root Mean Square Error (FRMSE), given as below:

$$FRMSE = \sqrt[4]{\frac{\sum_{i=1}^{m}(t_i - t_i^*)^2}{m}}, \qquad (2)$$

where $t_i$ is output of SONFIS and $t_i^*$ is real answer; $m$ is the number of test data (test objects). Figures 3 b indicates the results of the aforesaid system in which 135 initial clusters take a balance state with two rules. Figure 3c shows our mean about the structure detection, where the colored arrows shows possible dominant structures on the data field. With more neurons in SOM, we can acquire like patterns but may lose the supposed balancing criteria. This proves the balance measure even if gets min RMSE but losses the data distributions or major structures (compare figure3.c and 4.b). A similar process by increasing of the rules to8 was repeated. In this case 15 granules are in balance with 8 rules (figure 3, 4). Despite the obtained error is lower than the former case, but the revealed structures in the first layer have been truncated and didn't cover all of the possible structures as well as disclosed in the first situation. Comparison of the left side attributes with the possibility distribution (or membership functions) proves that the pick of former distribution is concurred with the density of degree functions in the latter distribution (Figure 5). Based on the latter case, we can plot the second granules 'relations.

Figure 6.a indicates that in the lower depth of the wells, increasing of solid content in the mud causing increase in mud loss rate.

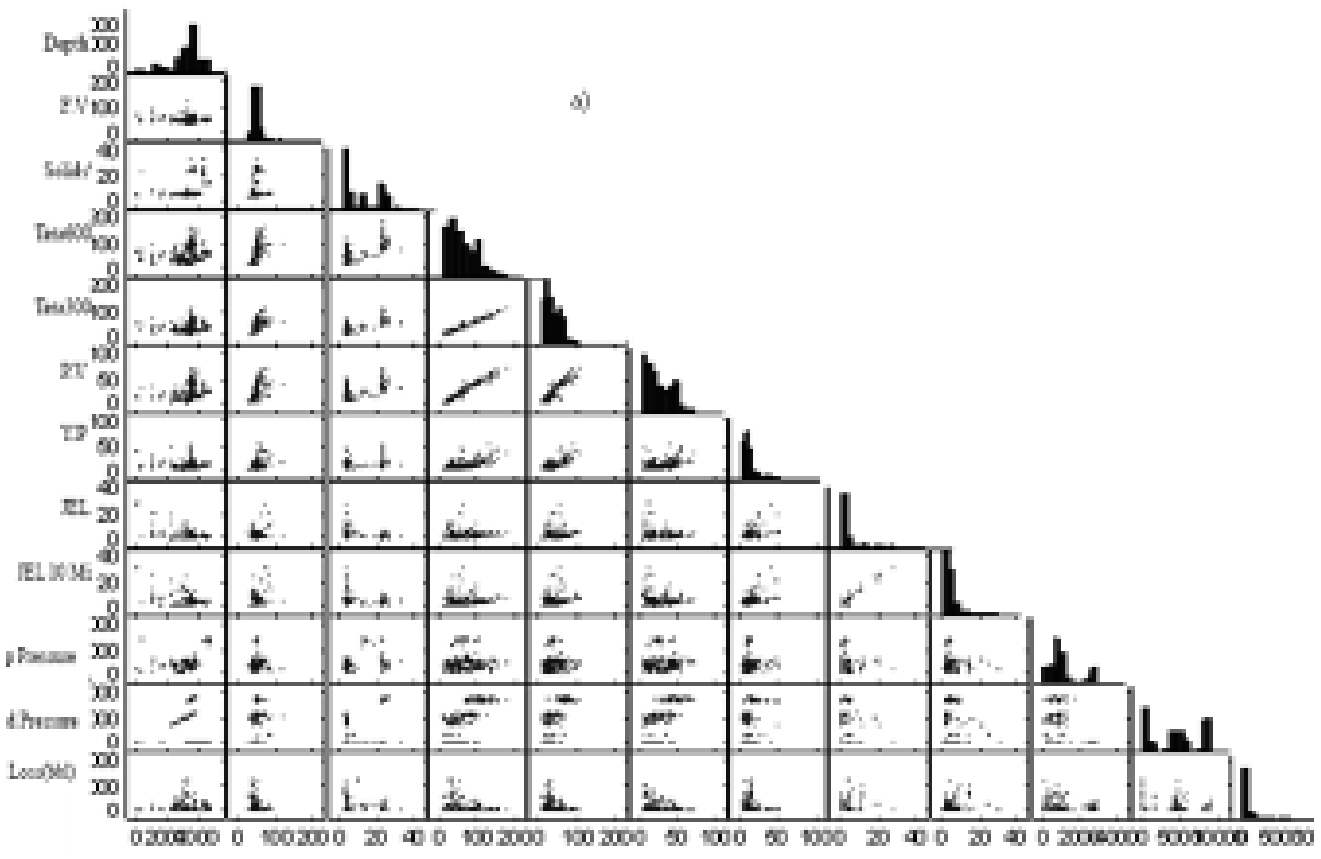

MW: Mud Weight, F.V: Funnel viscosity, Solid: the percent of solids in the Mud, P.V: plastic Viscosity, Y.P: Yield Point of Mud, RPM loss (BBL): the rate of Mud losing &P.P; Pump Pressure

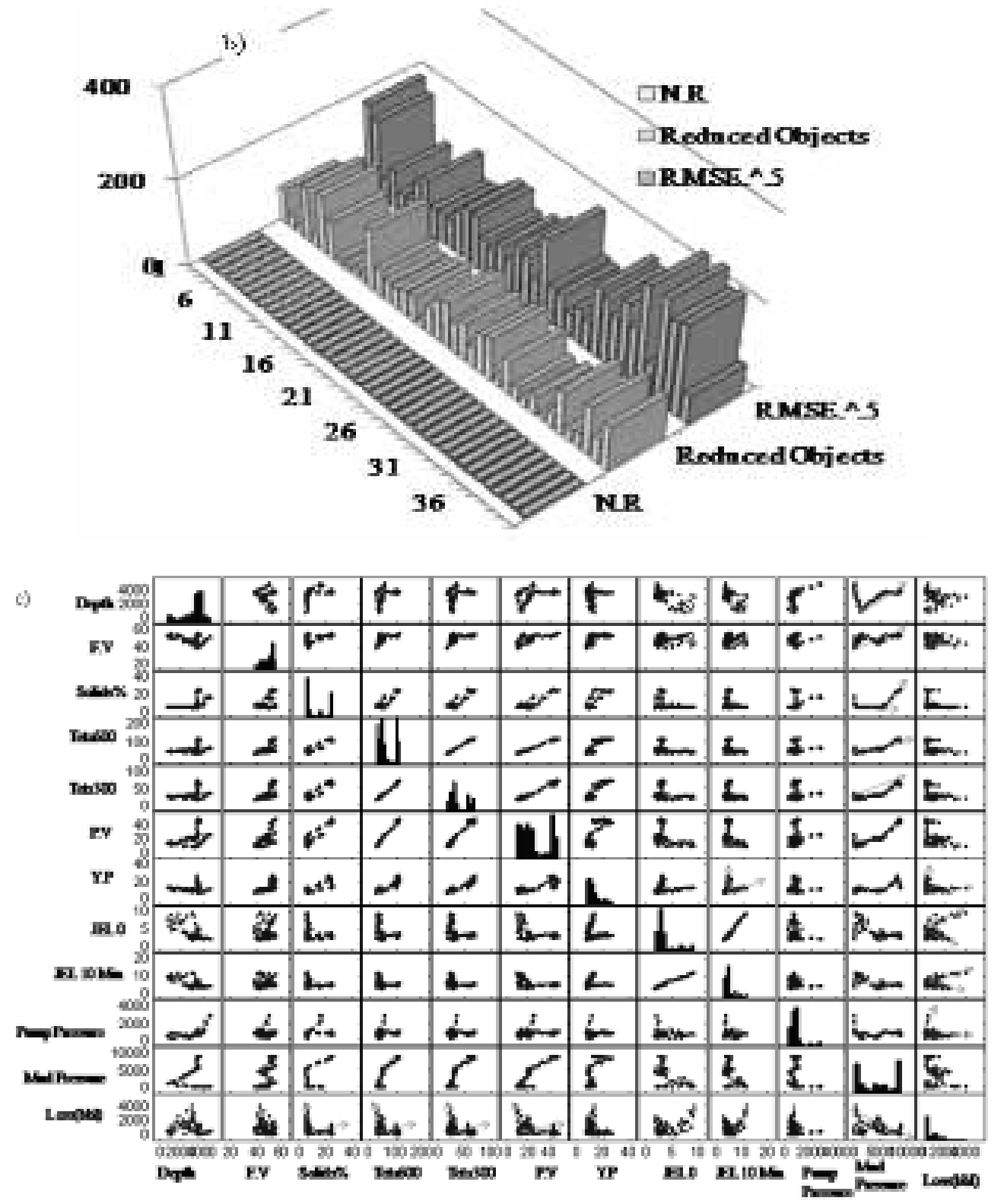

Fig.3. a) Initial information table with 11 condition attributes and Mud loss as a decision part ;b) SONFIS-R results with maximum number of rules =4 and close-open iterations = 10 and c) A reduced form of crisp (non-fuzzy) granules ( 27* 5 neurons in SOM) in balance with 2 rules obtained by SONFIS

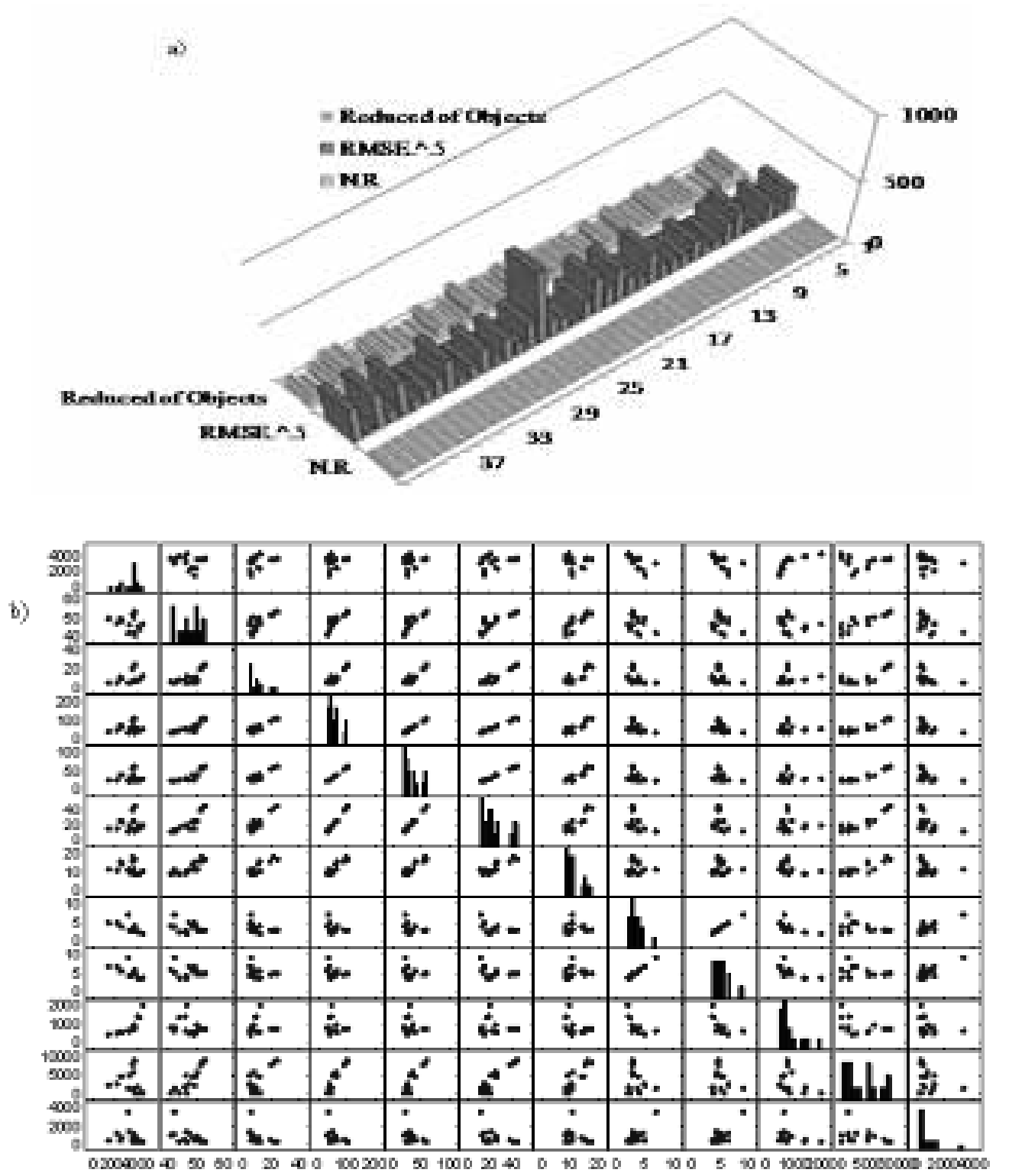

Fig.4. a) SONFIS-R results with maximum number of rules =8 (min rules :5) and close-open iterations = 10 and c) A reduced form of crisp (non-fuzzy) granules ( 3* 5 neurons in SOM) in balance with 8 rules obtained by SONFIS

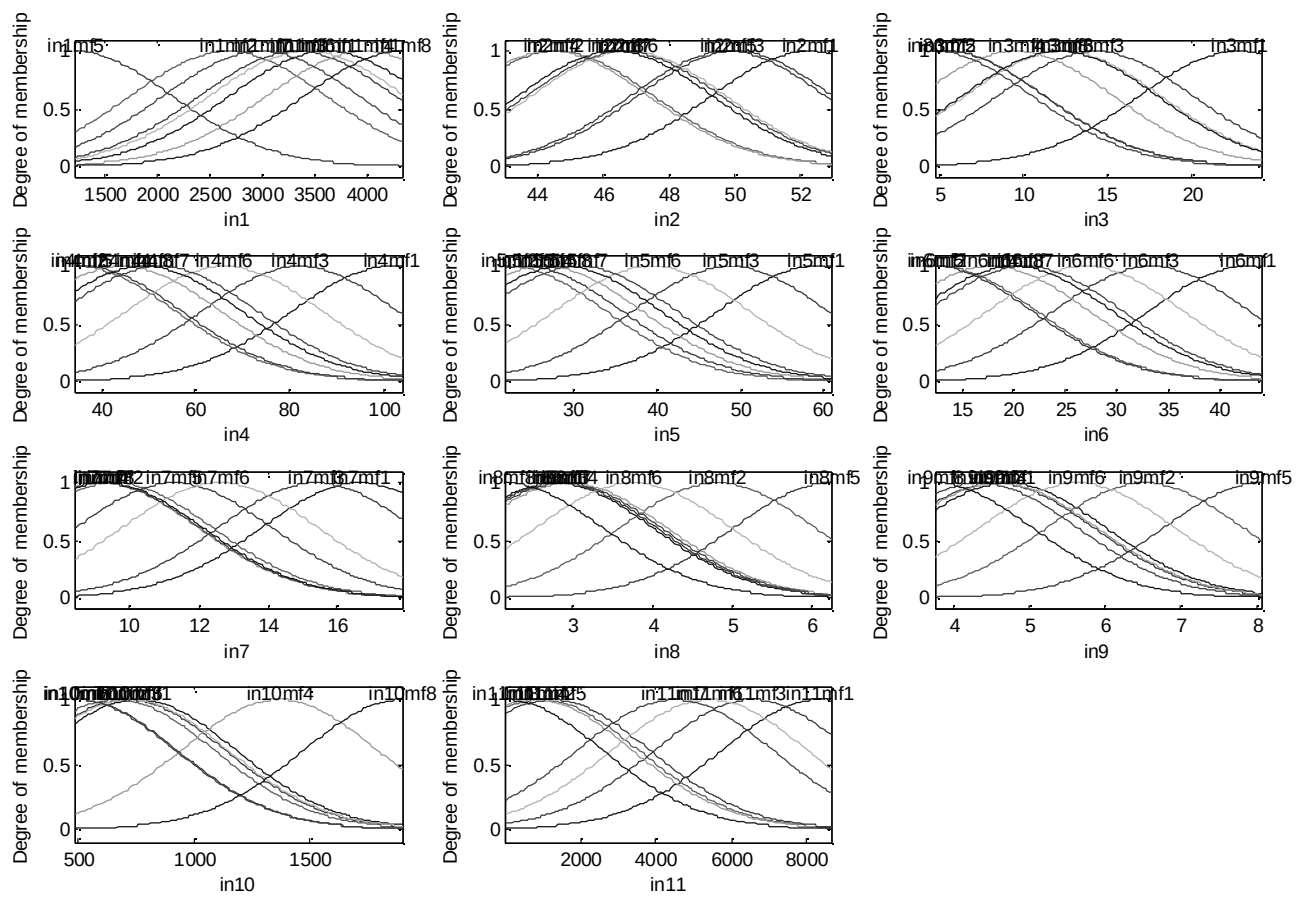

Fig.5. Best possibility distribution (with 8 membership's functions) of the conditional attributes (inputs) in the balance with 15 crisp granules

It seems that the reason of this event is using low density mud with low rheological properties, therefore sensitivity to the solid content is high and increasing in the solid content is one of the main factors that control the mud loss rate in the shallow depth. To overcome this problem accurate monitoring of solid content and utilization of solid control equipment in mud cycle is essential.

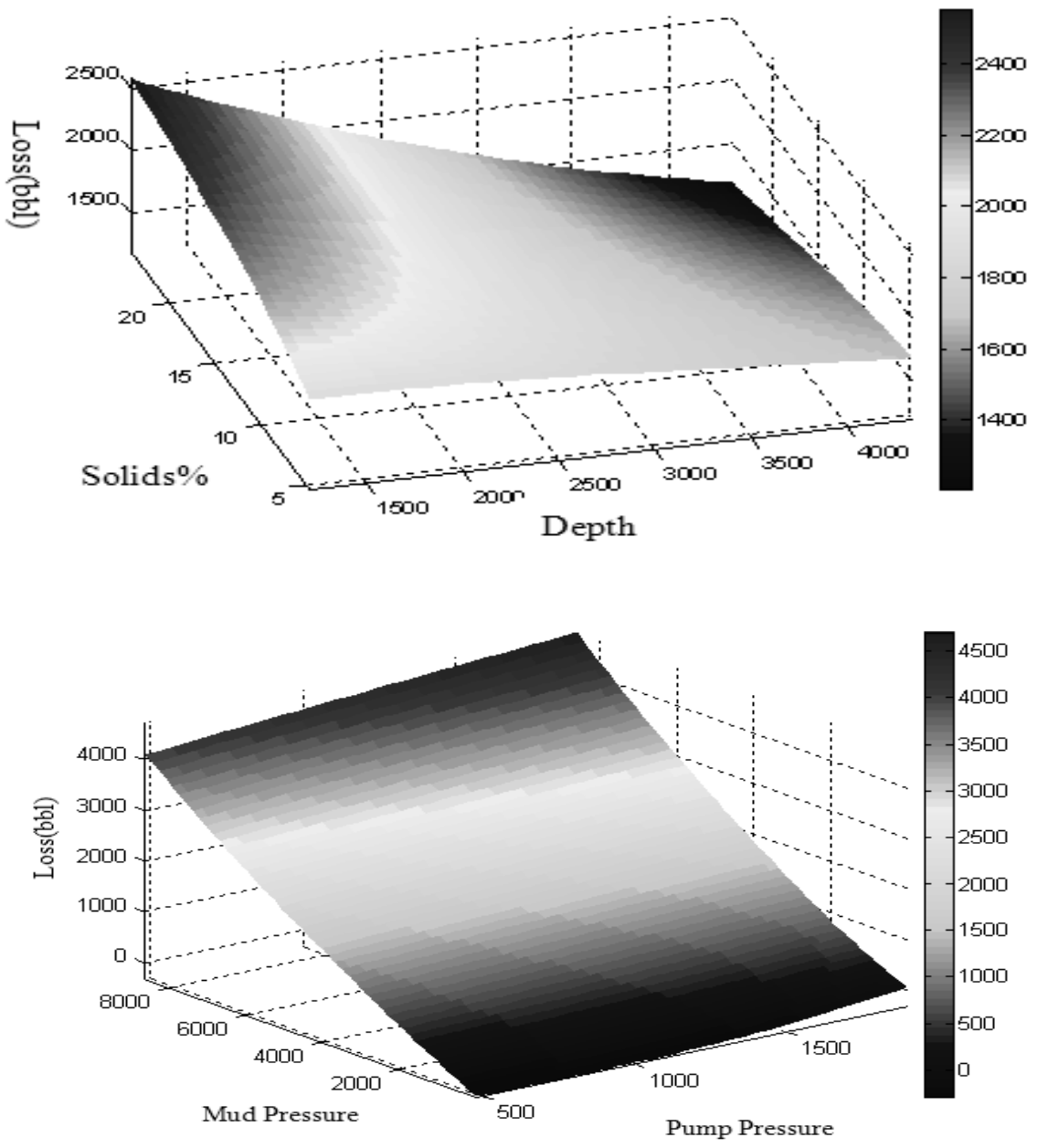

Fig.6. 3-D plots between selected attributes using SONFIS with 8 rules

Figure 6.b shows that sever lost circulations occur in deep intervals, at such depths increasing of PV, YP and other rheological properties of mud is the major reason of the sever lost circulations, hence maintaining mud properties in planed limits when using high density mud in deep intervals is useful for solving of lost circulation problem. These results are coincided with the revealed structures in figure 3.c or figure 4.b.

## 2.2. Large Scale Permeability using SORST

In this part of our study, we pursue a practical example, which covers a comprehensive data set from lugeon test in Shivashan dam. Shivashan hydroelectric earth dam is located 45km north of Sardasht city in northwestern of Iran.

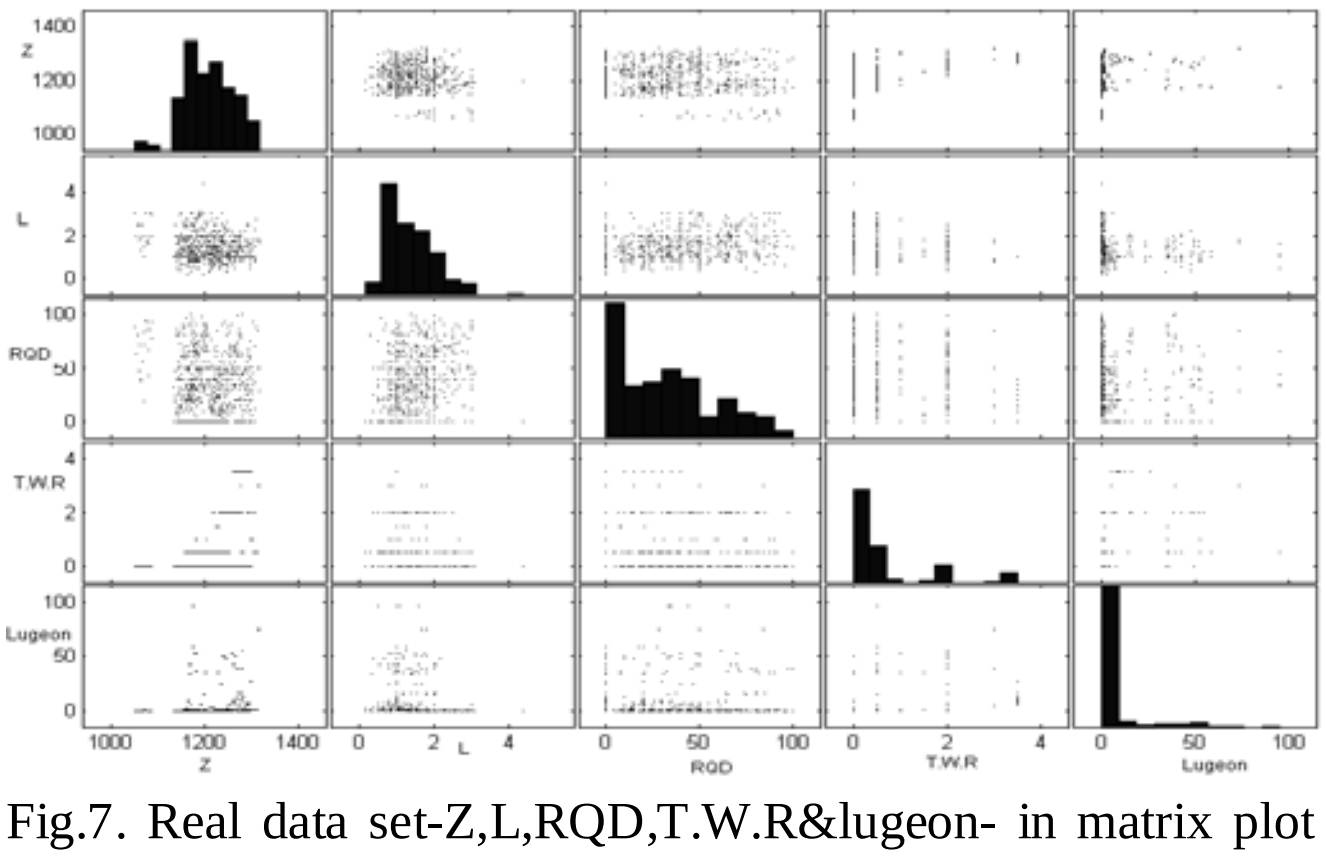

Fig.7. Real data set-Z,L,RQD,T.W.R&lugeon- in matrix plot form

The width of the V-shaped valley with similarly sloping flanks, at the elevation of 1185*m* and 1310*m* with respect to sea level are 38*m* and 467*m*, respectively. A general pattern from five chief attributes of the boreholes, consist of *Z*: elevation of the test, L: length of the tested section, RQD and Type of Weathering Rock (*T.W.R*) has been shown in figure7.

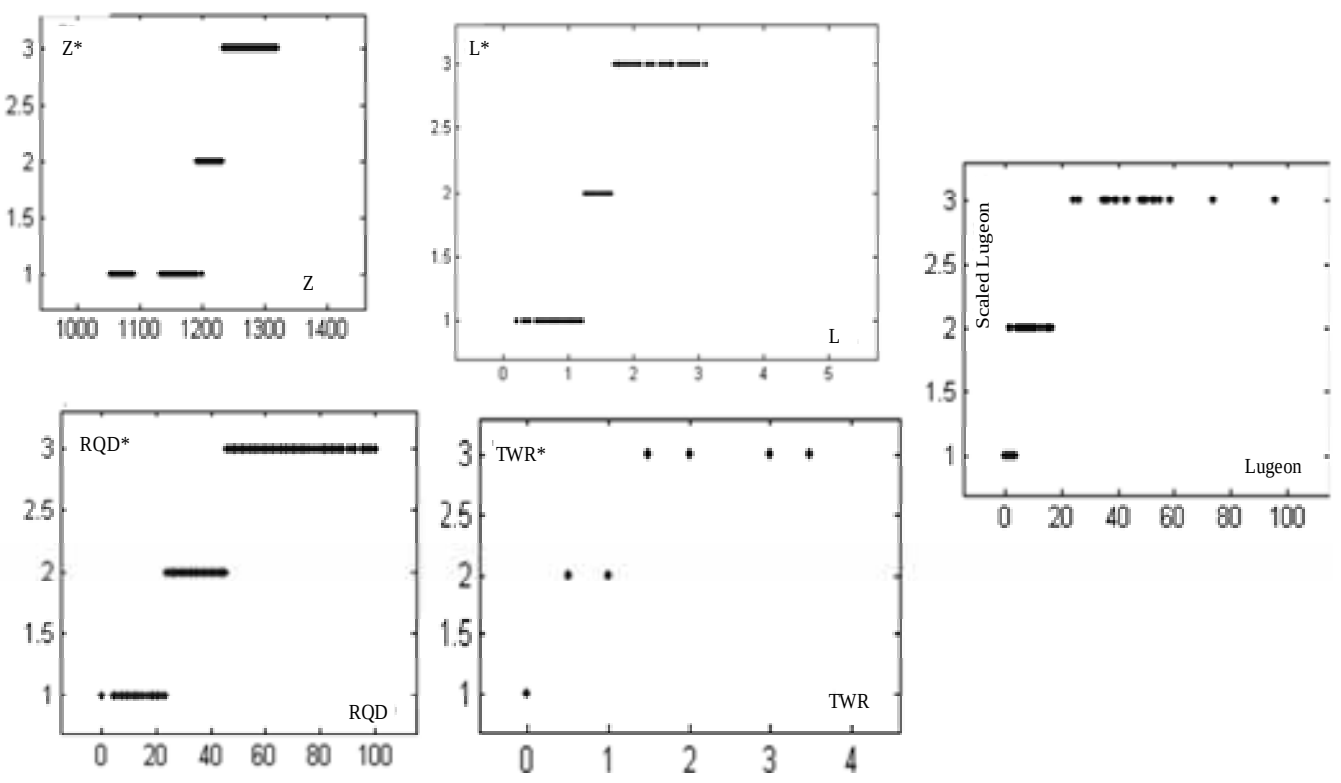

Fig 8. Results of transferring attribute (Z, L, RQD, T.W.R, and lugeon) in three categories (vertical axes) by 1-D SOM

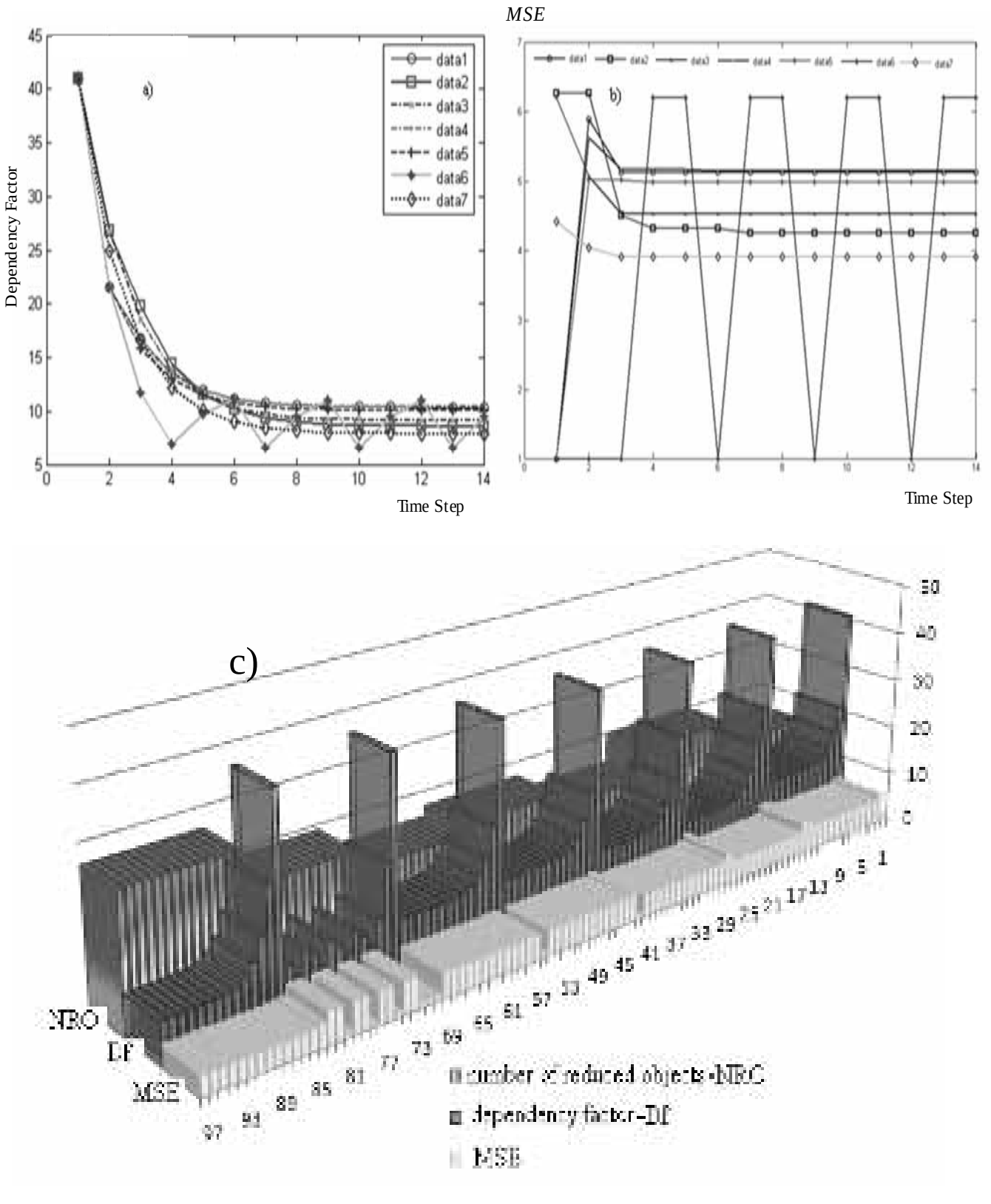

Fig.9. SORST-R results on the lugeon data set: a) strength factor; b) error measure variations along strength factor updating and c) 3-D column perspective of error measure, neuron changes and dependency factor (strength)

Using SORST and with exact rules i.e., one decision class in right hand of an *if-then* rule, the approximated rules will be elicited. Figures 8 and 9 depict the scaling process by 1-D SOM (3 neurons) and the performance of SORST-R over 7 random selection of SOM structure, respectively. The applied error measure is (Mean Square Error) :

$$MSE = \frac{\sum_{i=1}^{m} (d_i^{real} - d_i^{classified})^2}{m}, \quad (3)$$

where, $d_i^{real}$ is real transferred decision (lugeon) and $d_i^{classified}$ is classified (answer of system) decision.

It must be noticed that for unrecognizable objects in test data (elicited by rules) a fix value such 4 is assigned. So for measure part when any object is not identified, 1 is attributed. This is main reason of such swing of MSE in reduced data set 6 (figure 9-b). Clearly, in data set 7 SORST gains a lowest error (26 neurons in SOM). The extruded rules in the optimum case can be perused in the table 1.

Table1. Rules on N=26 selected among 696 objects;by SORST-R {Type of Weathering Rock (TWR); Dec: Decision (scaled lugeon value: 1: Low, 2: Medium, & 3: High); l: Length of Section; Z: Elevation}

| | |
|---|---|
| 1 | (Z = 2) => (Dec = 1); |
| 2 | (l in {2, 3}) & (RQD = 2) => (Dec = 1); |
| 3 | (Z = 3) & (l = 2) & (RQD = 1) => (Dec = 3); |
| 4 | (l = 2) & (TWR = 3) => (Dec = 3); |
| 5 | (Z = 3) & (l = 1) => (Dec = 1) OR (Dec = 3); |
| 6 | (l in {1, 2}) & (TWR = 2) => (Dec = 2); |
| 7 | (RQD = 2) & (TWR = 3) => (Dec = 2) OR (Dec = 3); |
| 8 | (Z = 1) & (RQD = 1) => (Dec = 2); |

## 3. A COMPLEX NETWORK APPROACH ON THE APERTURE EVOLVING

A network (graph) consists of nodes and edges connecting them. In a stochastic network, connection is probabilistic. A connectivity distribution (or degree distribution), P(*k*), is the probability of finding nodes with *k* edges in Network. Connectivity of a random graph obeys the Poissonian distribution in the case that the numbers of nodes are limited [11]. Complex networks have been developed in the several fields of science and engineering for example social, information, technological, biological and earthquake networks [12], [13]. In this section, the aperture evolving of a rock joint under gradual increasing of the shear displacement based on a complex network is evaluated. The rock material was granite with the weight of 25.9 $KN/m^3$ and uniaxial compressive strength of 172 Mpa. An artificial rock joint was made at mid height of the specimen by splitting using special joint creating apparatus, which has two horizontal jacks and a vertical jack [14],[15]. The sides of the joint are cut down after creating joint and its final size is 180 mm in length, 100 mm in width and 80 mm in height. Using special mechanical units the different mechanical parameters of this sample were measured. A virtual mesh having a square element size of 0.2 mm spread on each surface and each position height was measured by the laser scanner. The details of the procedure can be followed in [15], [16]. In this study our aim is to utilizing of a network approach on the rock joint surface so that characterization of the appropriate aperture (figure 10) behavior under successive shear displacements is accomplished by the appropriate network. To covering a network on the aperture patterns, we consider each X-profile (aperture profiles parallel with the Y-axis) as a node. To make edge between two nodes, the Euclidian distance:

$$d = \sqrt{\sum_{i,j=1:n_{px}} (\mathbf{x}_i(x_1, x_2, ..., x_n) - \mathbf{x}_j(x_1, x_2, ..., x_n))^2}, \quad (4)$$

is used where $\mathbf{x}_i$ is the $i^{th}$ profile and $x_k^{\,i}$ shows $k^{th}$ element from the $i^{th}$ profile. When $d \le \xi$ an edge among $i^{th}$ and $j^{th}$ profile will be created. The threshold $\xi$ depicts error level (Here we put $\xi$=5). In this way, the similarity of the aperture profiles is assessed.

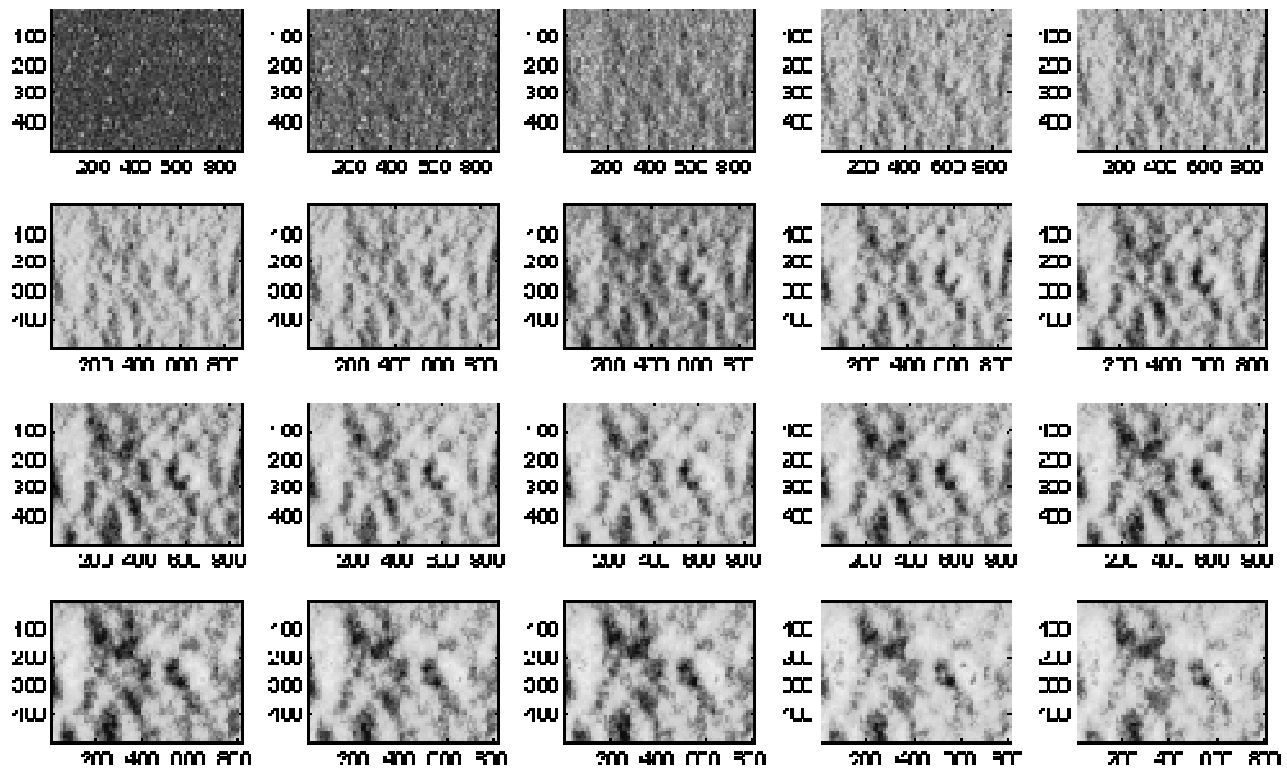

Fig.10. Aperture evolving under shear displacement (1-20 *mm*)

Figures 11, 12 illustrate the matrix form of the related network on the aperture state at 3*mm* shear displacement and the appropriate visualized network (only first 100 nodes among 801 nodes) at 20 mm displacement, respectively. The clusters of the similar aperture profiles in figure 11 (dark colors indicates the being of the edge) confirms that each profile is similar with its neighborhoods. Characterization of these similarities under the complex network approach show like behaviors in the systems having phase transition step(s) as well as the emerged manner in the dilation (figure 14.d).

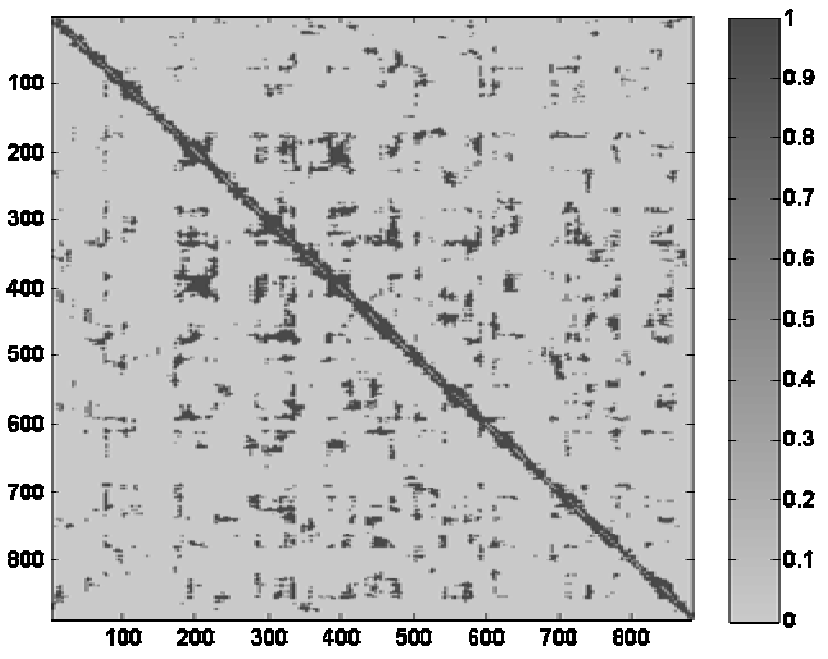

Fig.11. matrix form of the assigned network on the x-profiles of the aperture at 3 *mm* shear displacement

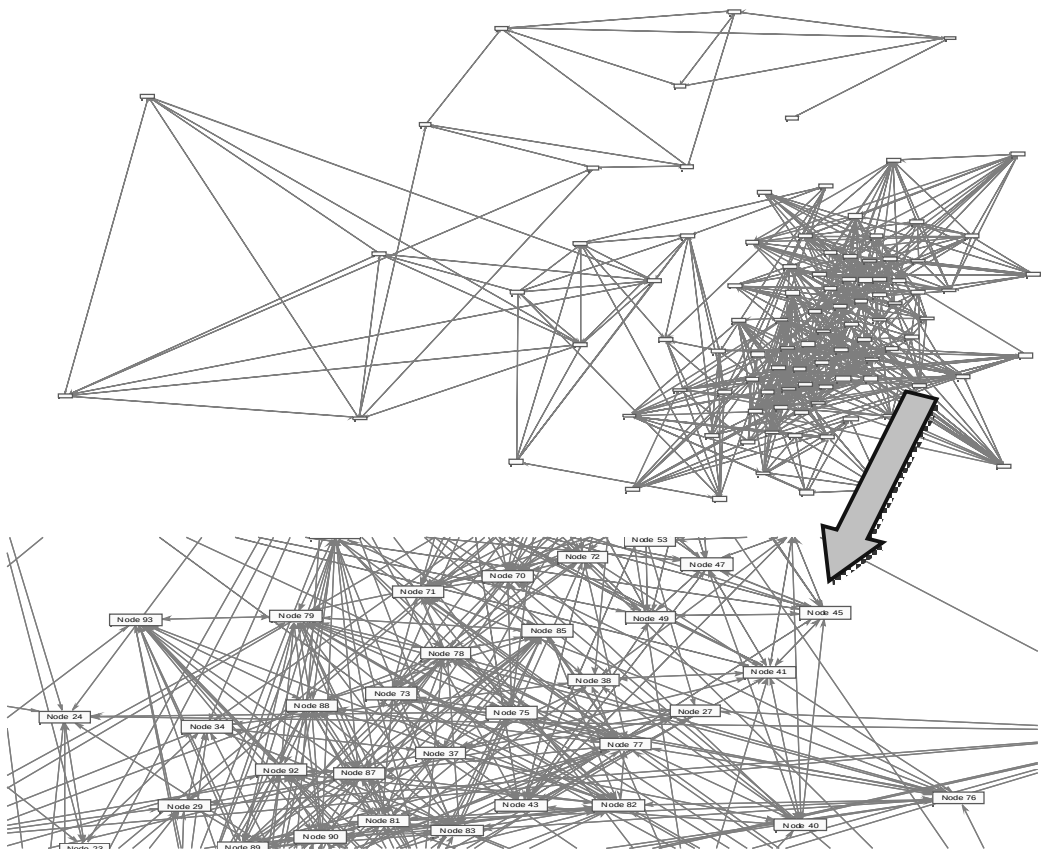

Fig.12. Part of the created network (only first 100 nodes) in 20 *mm* shear displacement

Regard to the P(*k*) evolving of the successive networks (figure 13) , distribution of *k* shows a soft transition from the semi-fully connected graph to the network with the similar Gaussian distribution. Such transition to the Gaussian state can be followed as a preferential removing of nodes (also edges). This preferentiality is resulted from our definition of the prevailed relate to make the networks. To distinguish the phase transforming step, where the asperities of the two surfaces are out of natural order, the evolution of the edges numbers versus shear displacement was plotted (figure 14.a ,b).

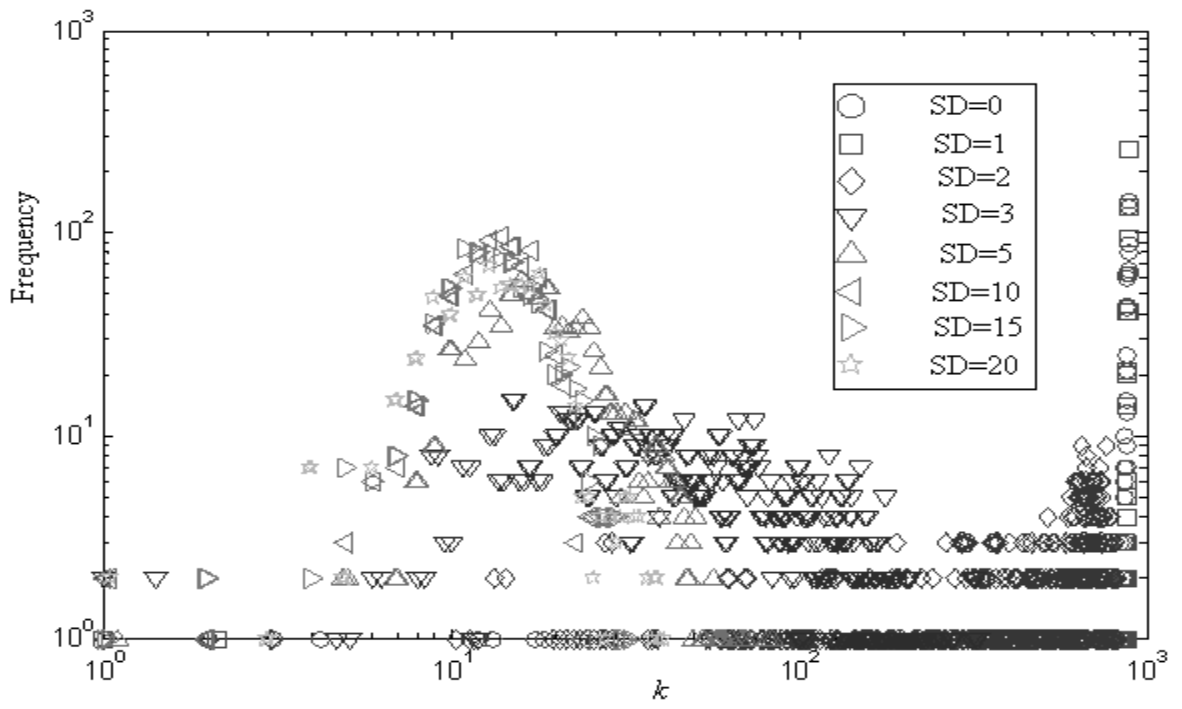

Fig.13. Evolution of the number of edge's frequency ( $\propto$ P(k)) during shear displacements in the log-log coordination

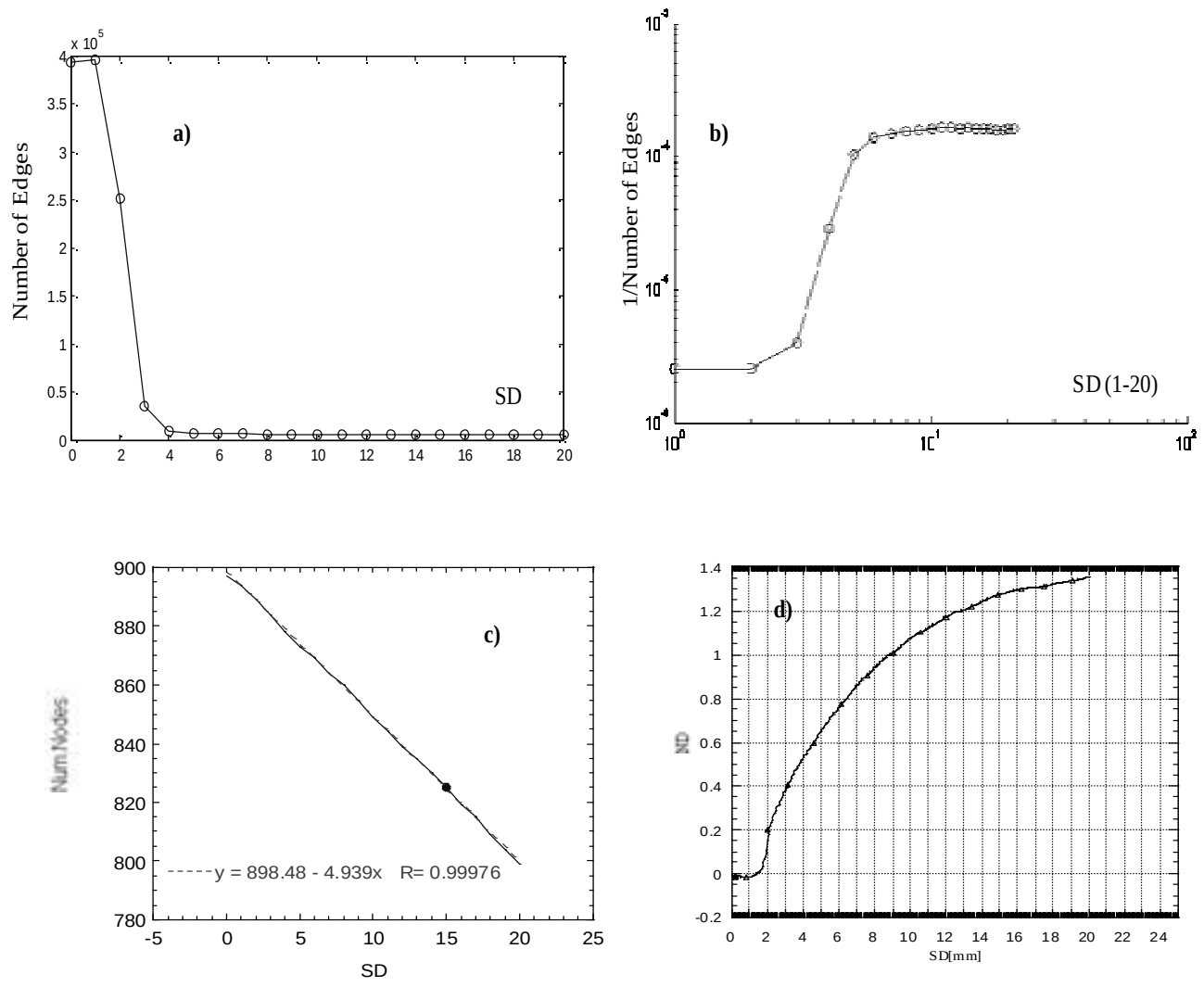

Fig.14. a) Number of Edges -Shear displacement, b) Log-Log diagram of 1/Number Edges-Shear displacement, c) Number of nodes -Shear displacement- and d) Joint normal displacement-Shear displacement

This feature is a guideline to assigning phase transition models to such networks as well as one may investigate the effect of the heterogeneity and anisotropy (x-profiles to the y-profiles) networks on the transition step [17], [18]. It must be noticed that the phase alteration in this case is not an order to disorder (or reveres case) but is a soft transition from an order to the semi-order case where after 4 *mm* the system takes a semi-stable state. We can nominate a sigmoid function for figure 14.b, is given as below [19]:

$$f(x) = [1 + \exp(\beta g(x) / g(x_0))]^{-1}$$
$$g(x) = -\ln(1 + \frac{x}{1 + \delta . x}) \qquad (5)$$

where *x is log(SD)* and *f(x)* shows *log(1/<k>).*

## 4. CONCLUSION

In this study, to developing of soft granules construction in not 1:1 mapping level of modeling, we proposed two main algorithms: Self Organizing Neuro-Fuzzy Inference System (SONFIS) and Self Organizing Rough Set Theory (SORST). Some advantages of the algorithms are new computations on the reduced data set based on hierarchical balancing of successive granules, finding out best fuzzy and rough rules in balance with the reduced data, assessing of SOM performance, elicitation of relative dominant structures on the data space and bridging between hard and soft computing methods (such as intelligent back analysis). So, we used our systems to analysis of permeability in Shivashan dam site and mud loss circulation in three oil wells at Fars area. I n the other part, as another branch of indirect modeling, using complex networks the aperture evolution was analyzed. The results show that without measuring additional mechanical character, one can estimate the transition step of the rock joint at the successive shear displacements. Characterization of the appeared network using analytical (or numerical) methods which make a similar behavior alteration can be put as an indirect modeling of a rock joint.